\documentclass[10pt, conference]{IEEEtran}
\usepackage{cite}
\usepackage{amsmath,amssymb,amsfonts}
\usepackage{algorithmic}
\usepackage{graphicx}
\usepackage{textcomp}
\usepackage{xcolor}
\usepackage{comment}
\usepackage[nolist,nohyperlinks]{acronym}
\usepackage{multirow}
\usepackage{url}

\graphicspath{{figures/}{FR3_ISAC/figures/}}

\usepackage[font=footnotesize]{subcaption}
\usepackage[font=footnotesize]{caption}

\newif\ifaccepted
\acceptedtrue 

\newcommand\inputrows[1]{\csname @@input\endcsname #1 }
\begin{document}

\acrodef{DPD}{Digital Predistortion}
\acrodef{DUT}{Device-Under-Test}
\acrodefplural{DUT}{Devices-Under-Test}
\acrodef{MP}{Memory Polynomial}
\acrodef{GMP}{Generalized Memory Polynomial}
\acrodef{ISAC}{Integrated Sensing and Communication}
\acrodef{PA}{Power Amplifier}
\acrodef{FR3}{Frequency Range 3}
\acrodef{NMSE}{Normalized Mean Square Error}
\acrodef{OFDM}{Orthogonal Frequency Division Multiplexing}
\acrodef{EVM}{Error Vector Magnitude}
\acrodef{BW}{Bandwidth}
\acrodef{VNA}{Vector Network Analyser}
\acrodef{IQ}{In-Phase and Quadrature}
\acrodef{NR}{New Radio}
\acrodef{QAM}{Quadrature Amplitude Modulation}
\acrodef{RRC}{Root-Raised-Cosine}
\acrodef{3GPP}{3rd Generation Partnership Project}
\acrodef{ACLR}{Adjacent Channel Leakage Ratio}
\acrodef{PAPR}{Peak-to-Average Power Ratio}

\title{Comparative Analysis and Behavioral Modelling of Power Amplifiers in FR3}

\author{\IEEEauthorblockN{Cel Thys\IEEEauthorrefmark{1}, Rodney Martinez Alonso\IEEEauthorrefmark{1}\IEEEauthorrefmark{2}, Ali H. Alsarraf\IEEEauthorrefmark{1},\\ Zhuangzhuang Cui\IEEEauthorrefmark{1}\IEEEauthorrefmark{2}, Dominique Schreurs\IEEEauthorrefmark{1}\IEEEauthorrefmark{2}, Sofie Pollin\IEEEauthorrefmark{1}\IEEEauthorrefmark{2}}
\IEEEauthorblockA{\IEEEauthorrefmark{1}WaveCoRE, Department of Electrical Engineering, Katholieke Universiteit Leuven (KU Leuven), Belgium\\
\IEEEauthorrefmark{2}Interuniversity Microelectronics Centre (IMEC), Leuven, Belgium}
}
\maketitle

\begin{abstract}
Power amplifier (PA) nonlinearity is a critical transmitter impairment in the upper mid-band (7\,GHz to 24\,GHz), designated as \ac{FR3} in emerging 6G standardisation.
The 3GPP PA model baseline, TR~38.803 Annex~A, was derived from 2.1\,GHz and 28\,GHz measurements and has not been revisited yet for \ac{FR3} standards work.
This paper addresses that gap with characterisation measurements of multiple PA devices-under-test (DUTs), driven into gain compression in the 7\,GHz and 15\,GHz bands with 100\,MHz and 200\,MHz bandwidth.
Memoryless polynomial and \ac{GMP} behavioral models are extracted by least-squares regression and compared with 3GPP models spanning FR1 and FR2.
The GaN PA DUT is modelled best at a higher nonlinear order and shows markedly stronger phase distortion than the wideband PA.
The memory terms of the \ac{GMP} model measurably improve the \ac{NMSE} at all operating points.
An orthogonal frequency-division multiplexing (OFDM) ambiguity function analysis from the \ac{ISAC} literature, using the extracted models, illustrates how PA compression raises the sidelobe floor of the sensing response.
These results provide a reference for PA impairment modelling in \ac{FR3}, directly applicable to \ac{ISAC} system design. All model coefficients and measurement data are publicly available~\cite{dataset_306iim}.
\end{abstract}


\acresetall

\section{Introduction}\label{sec:introduction}
The sixth generation (6G) of wireless networks is expected to expand into the upper mid-band from 7 to 24\,GHz, designated as \ac{FR3} in \ac{3GPP} studies~\cite{3gpp_38820}. It combines coverage close to that of sub-6\,GHz systems with wider contiguous bandwidths, and is a candidate range for 6G deployments. The ITU-R has identified this range as a priority for future terrestrial networks~\cite{itur_m2160}, and the 7.125--8.4\,GHz and 14.8--15.35\,GHz windows within it are already under study for future allocation~\cite{wrc23_res256}.

\ac{ISAC}, proposed for 6G, lets the radio network support communication and radar sensing with a shared waveform~\cite{liu2022isac,3gpp_22837}. FR3 is a suitable candidate band for \ac{ISAC}~\cite{cui_6g_2025}. Its wide contiguous bandwidths give fine range resolution, its sub-bands at 7\,GHz and 15\,GHz allow dual-band sensing, and its propagation loss is lower than at millimetre-wave frequencies.

The \ac{PA} nonlinearity is the dominant source of transmitter distortion. In communication, nonlinearity degrades the in-band modulation signal quality and spreads power into out-of-band frequency channels. In sensing, the same distortion raises the range sidelobes of the ambiguity function and impairs target detection~\cite{feng2024ofdmradar}. Studies of PA-impaired \ac{ISAC} use a memoryless soft limiter~\cite{gourar2025pa_isac} or a generic 3rd-order polynomial~\cite{feng2024ofdmradar} rather than models of measured hardware. Link- and system-level simulation needs behavioral models that map the complex baseband input directly to the output and are fitted to measurements of actual PAs \cite{kim2001mp, morgan2006gmp}. The circuit-level models used in PA design describe the transistors and matching networks, a level of detail these simulations do not need. The only 3GPP baseline, TR~38.803 Annex~A~\cite{3gpp_38803}, was derived from measurements at 2.1\,GHz and from circuit simulations at 28\,GHz, and has not been validated by measurement in FR3.

To the best of our knowledge, no measurement-based PA behavioral model has been published for the 7\,GHz and 15\,GHz bands. This paper makes the following contributions:
\begin{itemize}
  \item We release a complex baseband measurement dataset for a GaN PA at 8\,GHz and a wideband PA at 7.225\,GHz and 15.0\,GHz with two signal bandwidths, so that others can fit their own models or run their own simulations~\cite{dataset_306iim}.
  \item We extract memoryless polynomial, \ac{MP} and \ac{GMP} models from our measurements and report the accuracy gain of including memory terms. We compare the models with the 3GPP baseline models and publish the polynomial coefficients with the dataset~\cite{dataset_306iim}.
  \item We evaluate the models in an OFDM \ac{ISAC} ambiguity-function simulation and quantify the sensing degradation caused by the PA nonlinearity.
\end{itemize}

\section{Methodology}\label{sec:system_model}
This section describes how the PAs are measured and modelled. It covers the measurement setup and the \acp{DUT}, the publicly released dataset and behavioral models. It also describes the metrics used for model selection and for comparison with the 3GPP reference~\cite{3gpp_38803}.

\subsection{Measurement Setup and DUTs}\label{sec:meas_setup}

\begin{figure}[htbp]
\includegraphics[width=\linewidth, keepaspectratio]{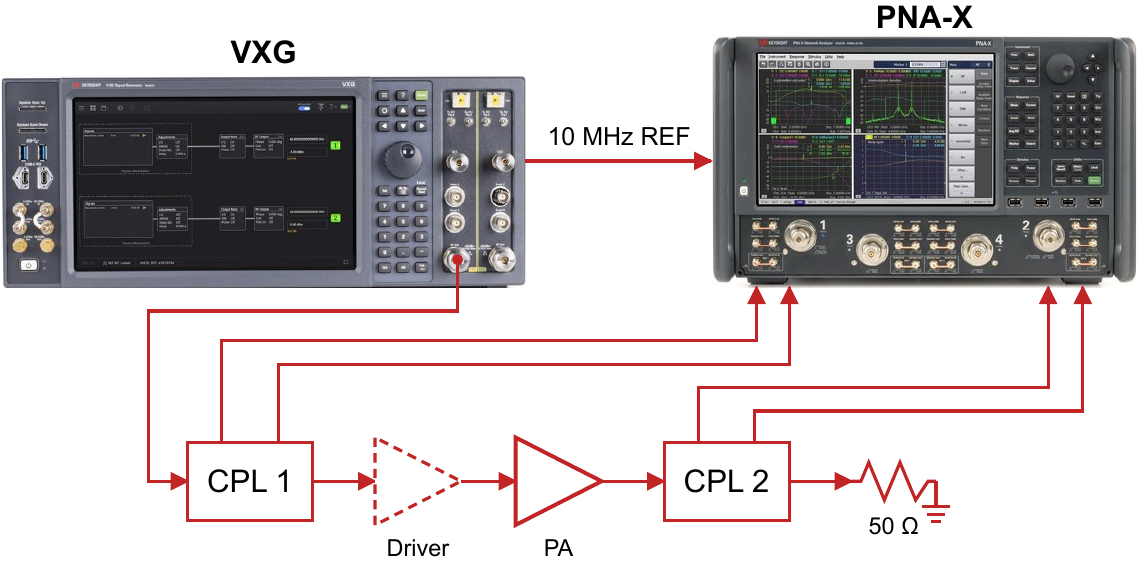}
\caption{\ac{VNA}-based PA characterisation setup. The linear driver is optional.}
\label{fig:meas_setup}
\vspace{-3mm}
\end{figure}

The setup in Fig.~\ref{fig:meas_setup} uses a Keysight M9484C VXG to generate the wideband waveforms and a PNA-X N5247B for calibrated acquisition~\cite{dunsmore_evolution_2024}. Before each measurement, a 2-port electronic calibration (Keysight N4691B ECal) places the reference planes at the \ac{DUT} connectors and corrects the cable losses, and a source calibration corrects the signal generation path. The VXG drives the \ac{DUT} through coupler CPL~1, and CPL~2 at the \ac{DUT} output ends in a 50\,$\Omega$ load. Both couplers feed the PNA-X receivers, which share a 10\,MHz reference with the VXG. Where needed, a linear driver amplifier (dashed in Fig.~\ref{fig:meas_setup}) precedes the \ac{DUT}, and a 30\,dB attenuator between the \ac{DUT} and CPL~2 (not drawn) protects the receiver. The attenuator is de-embedded in all reported gains and output powers. The drive level is set on the VXG, and refers to the RMS power of the modulated input signal. Our stimulus waveforms have a \ac{PAPR} of about 7dB, so each capture spans a range of instantaneous input powers, from the small-signal region up to the signal peak.

Three \acp{DUT} are characterised. The primary \ac{DUT} is a Qorvo TGA3042 GaN PA (7--10.5\,GHz, 4.5\,W) on its evaluation board (``GaN PA''). It is preceded by a driver, an Analog Devices HMC998APM5E GaAs amplifier with 15\,dB gain. The driver output stays more than 20\,dB below its 32\,dBm $P_\mathrm{1dB}$, so the measured nonlinearity is that of the GaN PA. It is measured at 8\,GHz, inside the 7.125--8.4\,GHz FR3 window~\cite{wrc23_res256}, with 100\,MHz \ac{BW} at 11 drive levels ($-25$ to $-15$\,dBm in 1\,dB steps). A Talent Microwave TLPA2G22G-43-43-HS wideband high-gain PA (``TLPA'') is measured at two frequencies and two bandwidths, 7.225\,GHz/200\,MHz and 15\,GHz with 100 and 200\,MHz \ac{BW}, which allows a comparison across bandwidth (Section~\ref{sec:bw_comparison}). An Analog Devices ADL9006 GaAs low-noise amplifier (``GaAs LNA''), driven at a much lower power level at 15\,GHz/100\,MHz, adds a comparison point for the 3GPP GaAs baseline (Section~\ref{sec:3gpp_comparison}). Both PAs use the 30\,dB output attenuator, and the GaAs LNA is connected directly.

We use a stimulus of 100\,MHz \ac{BW} for all characterisation measurements. This stimulus signal is single-carrier 64-\ac{QAM} sampled at 1280\,MHz, with 262\,144 samples per capture. For the TLPA \ac{DUT}, we also investigate a 200\,MHz stimulus signal, with 256-\ac{QAM} sampled at 2560\,MHz using 524\,288 samples. Both use \ac{RRC} pulse shaping with roll-off 0.35 at 80 and 160\,MBd. The $12.8\times$ oversampling captures the higher-order intermodulation products without risk of aliasing.

\subsection{Public Dataset}\label{sec:dataset}
The measurements are publicly available~\cite{dataset_306iim}: the 11 GaN PA drive levels, the three TLPA operating points and the GaAs LNA of Table~\ref{tab:model_params}, as well as four TLPA captures at 15\,GHz/100\,MHz with a swept input power. Each operating point has two or three independent captures with the same stimulus waveform, each holding the complex baseband input and output \ac{IQ} sequences $x(n)$ and $y(n)$. The sequences are time-aligned in MATLAB using cross-correlation to determine the integer and sub-sample delays, and a common phase rotation. 

This public dataset can be used to incorporate FR3 PA distortion modelling into simulations, exploring the impact on communication or sensing performance. It has also been used to support work on PA nonlinearity compensation using \ac{DPD} \cite{thys_low_2026}.

\subsection{PA Behavioral Model}\label{sec:pa_model}
The simplest model is a memoryless odd-order polynomial in the complex baseband:
\begin{equation}
  \hat{y}(n) = \sum_{k=0}^{K} a_{2k+1}\, x(n)\, |x(n)|^{2k},
  \label{eq:pa_poly}
\end{equation}
with coefficients $a_{2k+1} \in \mathbb{C}$ and order $2K+1$. The odd-order terms model the in-band intermodulation products of the complex envelope~\cite{pedro2003}. The coefficients are extracted by least squares fit:
\begin{equation}
  \hat{\mathbf{a}} = \bigl(\mathbf{\Phi}^H \mathbf{\Phi}\bigr)^{-1} \mathbf{\Phi}^H \mathbf{y},
  \label{eq:ls}
\end{equation}
where $\mathbf{\Phi} = \bigl[\mathbf{x},\;\mathbf{x}|\mathbf{x}|^{2},\;\ldots,\;\mathbf{x}|\mathbf{x}|^{2K}\bigr]$ is the $N\!\times\!(K{+}1)$ regression matrix.
The memoryless parameter $K$ is swept from $1,\ldots,6$ (nonlinear orders 3 to 13).
To capture memory effects, we fit a \ac{GMP} model~\cite{morgan2006gmp}, which offers a good accuracy/complexity tradeoff~\cite{tehrani2010tradeoff}, in the structure of the 3GPP \ac{GMP} models~\cite{3gpp_38803}:
\begin{equation}
\begin{split}
  \hat{y}_{GMP}(n) &= \sum_{l=-L}^{L} a_{0l}\, x(n{-}l) \\
             &+ \sum_{k=1}^{K}\sum_{l \in \mathcal{L}_A} a_{kl}\, x(n{-}l)\,|x(n{-}l)|^{2k} \\
             &+ \sum_{k=1}^{K}\sum_{l \in \mathcal{L}_B} b_{kl}\, x(n{-}l)\,|x(n)|^{2k},
\end{split}
  \label{eq:gmp}
\end{equation}
with non-causal lag sets $\mathcal{L}_A = \{-A,\ldots,A\}$ and $\mathcal{L}_B = \{\pm1,\ldots,\pm B\}$, $L=\max(A,B)$, and the cross-term envelope at the current sample ($m=0$ in~\cite{3gpp_38803}). With $B=0$, \eqref{eq:gmp} reduces to the \ac{MP}~\cite{kim2001mp}, and with $A=B=0$ to~\eqref{eq:pa_poly}. The \ac{MP} adds delayed copies of the memoryless basis functions of \eqref{eq:pa_poly}, so it models memory that acts on a past sample and its envelope together. The cross terms of the \ac{GMP} add the interaction of a past sample with the current envelope. The coefficients are also fit using least-squares~\eqref{eq:ls}, with a parameter search grid of $K=1,\ldots,6$, $A=0,\ldots,3$, $B=0,\ldots,4$. The model selection strategy is detailed next.

\subsection{Model Evaluation}\label{sec:model_eval}
Model quality is measured by the \ac{NMSE}
\begin{equation}
  \mathrm{NMSE} = 10\cdot\log_{10}\!\left(\frac{\sum_{n=1}^{N}|y(n)-\hat{y}(n)|^2}{\sum_{n=1}^{N}|y(n)|^2}\right)\,[\mathrm{dB}].
  \label{eq:nmse}
\end{equation}
Every model is fitted and selected on separate, independent parts of the data. We select the model with the fewest coefficients whose validation \ac{NMSE} is within 0.1\,dB of the lowest overall \ac{NMSE}. All reported \ac{NMSE} values are computed on a further test capture that is used neither for fitting nor for selection.

In what follows, we describe each \ac{PA} capture and/or model by its amplitude-to-amplitude (AM/AM) and amplitude-to-phase (AM/PM) conversion, i.e., its gain and phase shift versus input power. Since each \ac{DUT} has a vastly different gain and phase-shift, we normalise each to the small-signal values. On the vertical axes, the AM/AM shows the normalised gain, i.e., the gain $|y|/|x|$ in dB relative to the small-signal gain, and the AM/PM shows the phase shift $\angle y-\angle x$ relative to the small-signal phase. For a model, $y$ is replaced by $\hat{y}$. 

The four memoryless models of \ac{3GPP} TR~38.803 Annex~A~\cite{3gpp_38803} serve as references. They have the form~\eqref{eq:pa_poly} and surround FR3 from below and above: a GaAs ($K=7$) and a GaN ($K=4$) device at 2.1\,GHz (FR1), both based on measurements, and a CMOS ($K=7$) and a GaN ($K=5$) device at 28\,GHz (FR2), both from circuit simulations. Each also has a \ac{GMP} variant. 

The \ac{3GPP} models are normalised to unit small-signal gain on the input range $|x|\le1$ and have no absolute power scale. We therefore plot all following curves against the input back-off from each device's own 1\,dB compression point $P_\mathrm{1dB}$,
\begin{equation}
  \mathrm{IBO} = 20\log_{10}\bigl(|x|/|x_\mathrm{1dB}|\bigr) \quad [\mathrm{dB}],
  \label{eq:ibo}
\end{equation}
where $|x_\mathrm{1dB}|$ is the input amplitude at the measured $P_\mathrm{1dB}$. Every AM/AM curve thus starts at 0\,dB and passes through $-1$\,dB at 0\,dB IBO (the $P_\mathrm{1dB}$ point). Likewise, each AM/PM curve starts with a $0^\circ$ phase shift, so the \acp{DUT} can be compared by the shape of their compression and phase shift. 

The 1\,dB compression point $P_\mathrm{1dB}$ is found as the input power at which the measured gain first drops 1\,dB below the small-signal gain, or for the 3GPP models, at which the model gain drops 1\,dB below its small-signal gain. The small-signal gain and phase of a measurement are its coherent gain $\sum y(n)x^*(n)/\sum|x(n)|^2$ over the 5000 lowest-amplitude samples. We use the coherent gain which, unlike $\sum|y(n)|^2/\sum|x(n)|^2$, is not biased by additive output noise. The GaN PA is driven furthest into gain compression: at the highest drive level, its peaks reach 34.5\,dBm at 5.2\,dB compression, close to its saturated output power of 36.4\,dBm. The TLPA and the GaAs LNA are driven just beyond their $P_\mathrm{1dB}$, to 1.2--2.1\,dB compression.

\section{Measurement Results and Analysis}\label{sec:results}
This section applies the method of Section~\ref{sec:system_model} to the measured PAs. It first shows the GaN PA over its drive levels, then compares the extracted models with the 3GPP baselines, quantifies the memory effects and the bandwidth dependence, and evaluates the impact of the PA nonlinearity on sensing.

\begin{table*}[t]
  \caption{Extracted PA model overview: memoryless polynomial (ML) vs.\ \ac{GMP}, with polynomial order $2K{+}1$ and number of complex coefficients. For fitted model coefficients, see Section~\ref{sec:coefficients} and \cite{dataset_306iim}. The 3GPP reference values are as published in TR~38.803 Annex~A~\cite{3gpp_38803}.}
  \label{tab:model_params}
  \centering
  \begin{tabular}{llccccc}
    \hline
    \ac{DUT}/Model & $f_c$, \ac{BW} & Order ML/GMP & Coef.\ ML/GMP & NMSE$_\mathrm{ML}$ [dB] & NMSE$_\mathrm{GMP}$ [dB] & $\Delta$NMSE [dB] \\
    \hline
    \multicolumn{7}{l}{\textit{Measured (this work)}} \\
\inputrows{tables/tab_models_measured.tex}
    \hline
    \multicolumn{7}{l}{\textit{3GPP TR~38.803 Annex~A \cite{3gpp_38803}, published NMSE}} \\
    GaAs (meas.) & 2.1\,GHz, 40\,MHz (FR1) & 15 / 15 & 8 / 56 & $-31.5$ & $-38.1$ & $-6.6$ \\
    GaN (meas.) & 2.1\,GHz, 40\,MHz (FR1) & 9 / 9 & 5 / 35 & $-34.5$ & $-40.6$ & $-6.1$ \\
    CMOS (sim.) & 28\,GHz, 400\,MHz (FR2) & 15 / 15 & 8 / 40 & $-32.1$ & $-41.7$ & $-9.6$ \\
    GaN (sim.) & 28\,GHz, 400\,MHz (FR2) & 11 / 11 & 6 / 42 & $-33.4$ & $-41.1$ & $-7.7$ \\
    \hline
  \end{tabular}
\end{table*}

\subsection{GaN PA Drive-Level Sweep}\label{sec:sweep}
\begin{figure}[htbp]
  \centering
  \includegraphics[width=\linewidth]{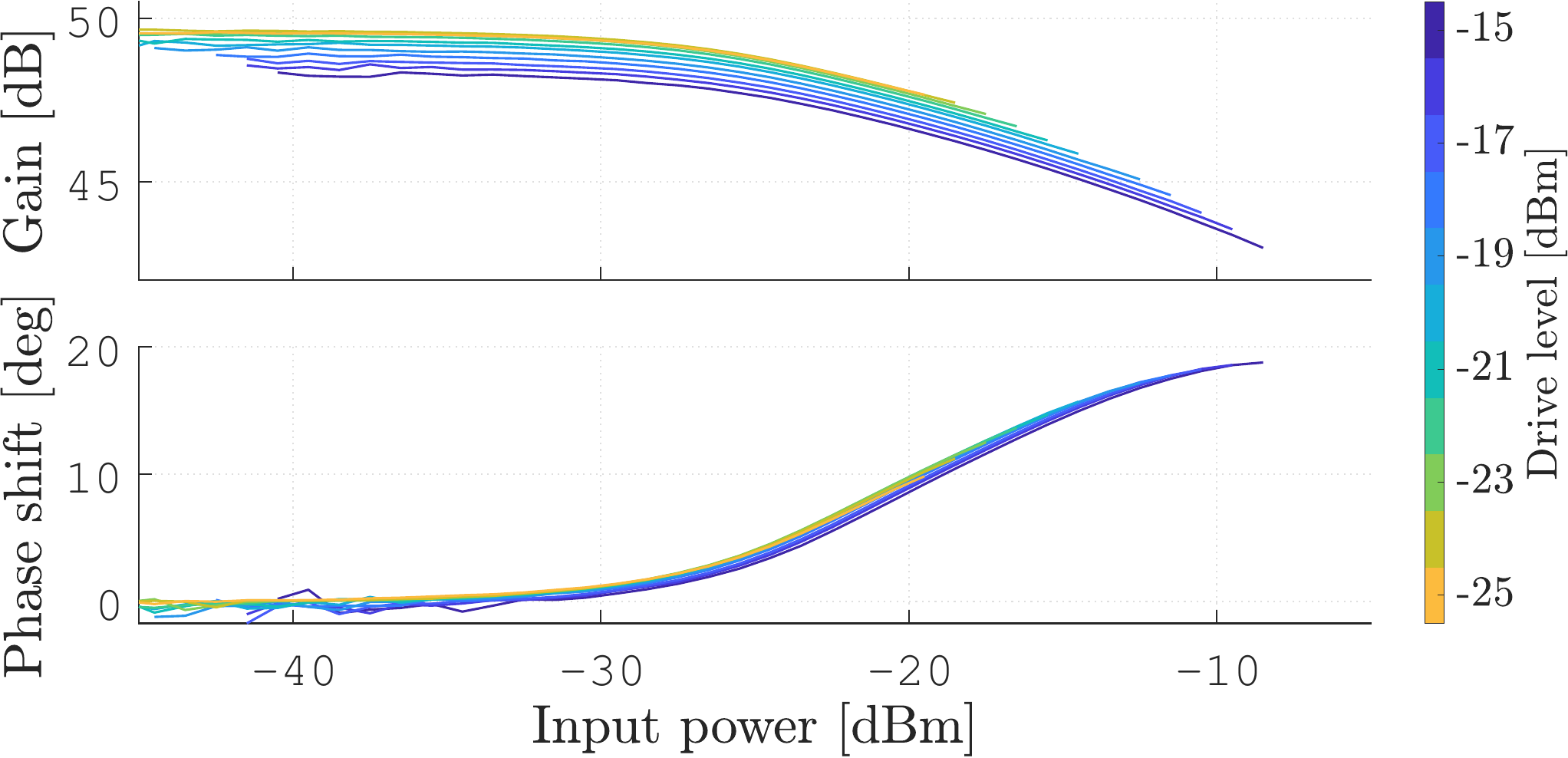}
  \caption{Measured gain (top) and phase shift (bottom) of the GaN PA at 8\,GHz, 100\,MHz for the 11 drive levels, vs.\ input power at the driver input.}
  \label{fig:tga_sweep}
  \vspace{-3mm}
\end{figure}

Fig.~\ref{fig:tga_sweep} shows the measured gain and phase of the GaN PA at its 11 drive levels, i.e., the magnitude and phase of the coherent gain (Section~\ref{sec:model_eval}) plotted against the power at the driver input. Across the 11 drive levels, the peak compression rises from 1.9 to 5.2\,dB, the output $P_\mathrm{1dB}$ stays at 24.5--25.0\,dBm, and the small-signal gain drops by 1.4\,dB with drive, so the levels do not lie on one AM/AM curve. The compression of the GaN PA is soft: the output $P_\mathrm{1dB}$ lies about 11\,dB below the datasheet saturated output, and in between both levels the gain slowly deteriorates. Thus, driving this \ac{DUT} at an efficient operating point would mean driving it beyond 1\,dB compression.

The selected nonlinear order of memoryless models ranges from 13 at the highest level to 5 at the lowest. In the rest of this section, we use the $-20$\,dBm drive level, which is about 3\,dB above $P_\mathrm{1dB,in}$, with 3.3\,dB peak compression. This is selected as a tradeoff between available power and nonlinear distortion. Its selected nonlinear order is 11, which means the memoryless model has 6 terms.

\subsection{Comparison with 3GPP Baselines}\label{sec:3gpp_comparison}
We compare the selected memoryless models (Section~\ref{sec:proposed_models}) with the four 3GPP baselines~\cite{3gpp_38803}. Our curves span the measured input range, which extends 1.9--5.1\,dB (TLPA) and 9.7\,dB (GaN PA) beyond the measured $P_\mathrm{1dB}$.

\begin{figure}[htbp]
  \centering
  \includegraphics[width=\linewidth]{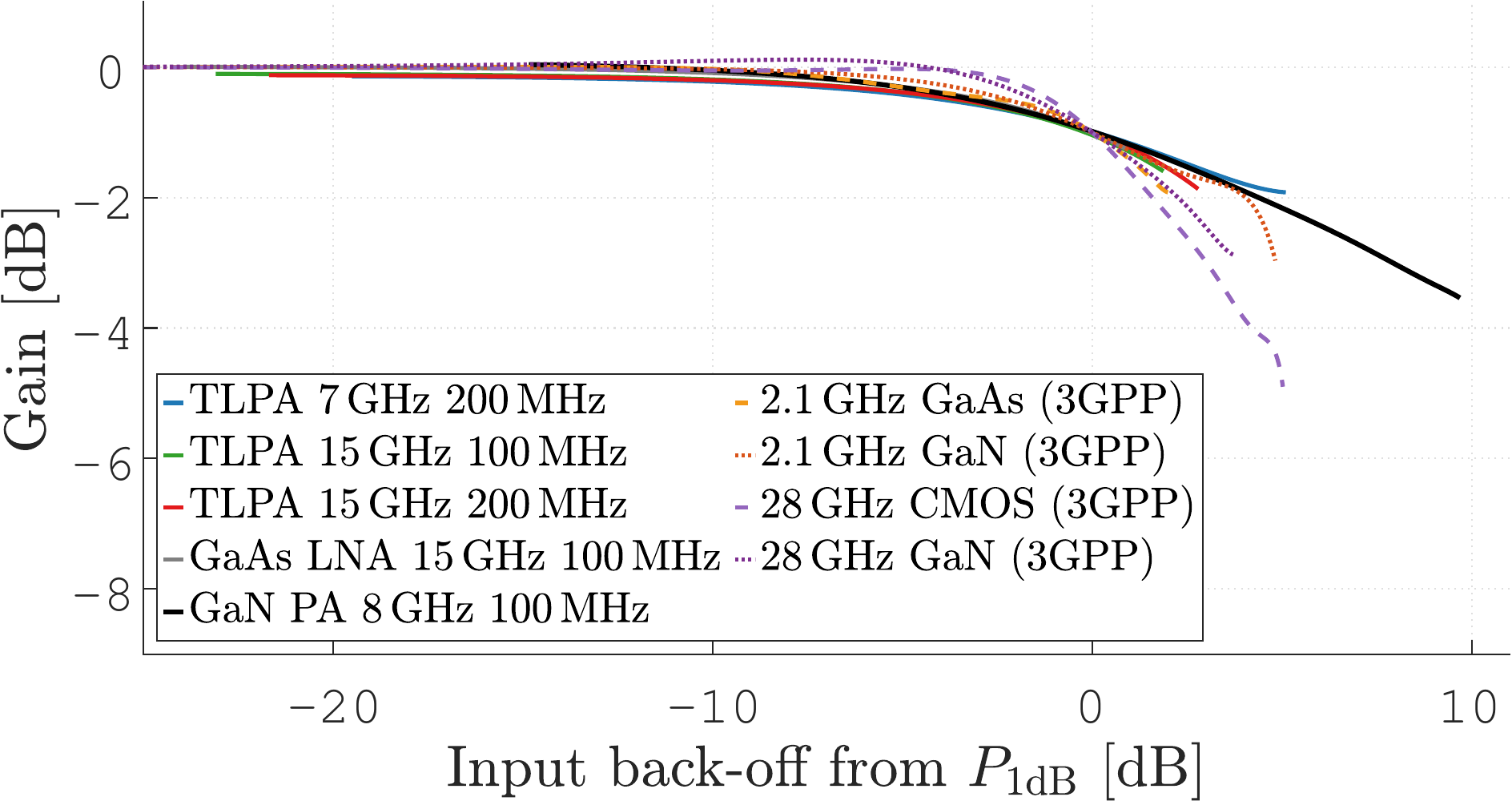}
  \caption{Normalised AM/AM vs.\ input back-off from $P_\mathrm{1dB}$: measurement-based models in FR3 vs.\ TR~38.803 Annex~A polynomial models. Gain is normalised by the small-signal gain of each \ac{DUT}.}
  \label{fig:amam_3gpp}
  \vspace{-3mm}
\end{figure}

\begin{figure}[htbp]
  \centering
  \includegraphics[width=\linewidth]{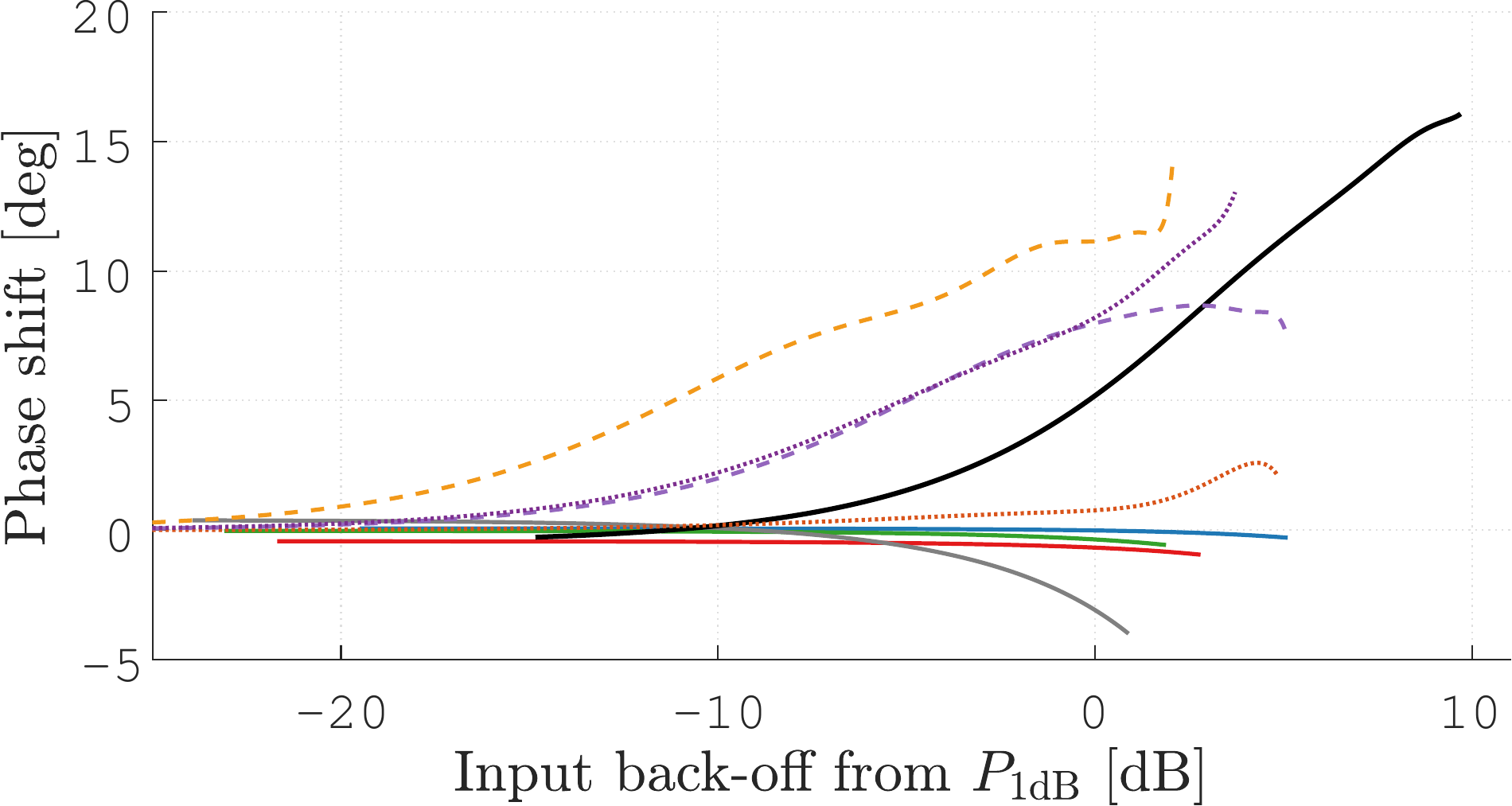}
  \caption{AM/PM vs.\ input back-off from $P_\mathrm{1dB}$: measurement-based models in FR3 vs.\ TR~38.803 Annex~A polynomial models. Legend is shared with Fig.~\ref{fig:amam_3gpp}. Phase shift relative to the small-signal phase shift of each \ac{DUT}.}
  \label{fig:ampm_3gpp}
  \vspace{-3mm}
\end{figure}

Fig.~\ref{fig:amam_3gpp} shows that up to $P_\mathrm{1dB}$, all models compress similarly, with less gain expansion in the extracted models than in the 3GPP baselines. Beyond $P_\mathrm{1dB}$, the 28\,GHz models compress fastest, while the TLPA and the GaN PA follow the 2.1\,GHz GaN baseline up to about 4\,dB beyond $P_\mathrm{1dB}$. The GaN PA then keeps compressing smoothly, to $-3.5$\,dB at 9.7\,dB beyond $P_\mathrm{1dB}$, a region the 3GPP models do not cover.

Fig.~\ref{fig:ampm_3gpp} shows that the phase shift differs more between the devices than the gain. Among the 3GPP models, the 2.1\,GHz GaAs model reaches $+14.3^\circ$ and the 28\,GHz GaN model $+13.0^\circ$, while the 2.1\,GHz GaN and 28\,GHz CMOS models stay at $+2.6^\circ$ and $+8.7^\circ$. The GaN PA reaches $+5.2^\circ$ at $P_\mathrm{1dB}$ and $+16.1^\circ$ at the edge of its measured range, more than any 3GPP reference, although at a larger IBO than these cover. The GaAs LNA, of the same technology as the 2.1\,GHz GaAs baseline, shows a phase shift of the opposite sign, about $-4^\circ$ instead of $+14.3^\circ$. The three TLPA operating points stay within $1^\circ$.

\subsection{Extracted PA Models and Memory Effects}\label{sec:proposed_models}\label{sec:gmp_comparison}
Table~\ref{tab:model_params} gives an overview of the selected memoryless and \ac{GMP} models of all measured operating points next to the four 3GPP references.
For the TLPA at 15\,GHz and the GaAs LNA, the validation \ac{NMSE} of the memoryless polynomial does not improve beyond 3rd order. The TLPA at 7.225\,GHz, with 2.1\,dB peak compression, needs 5th order. The strongly compressed GaN PA needs 11th order: its test \ac{NMSE} improves from $-29.4$\,dB at 3rd order to $-39.4$\,dB at 11th order. Section~\ref{sec:coefficients} lists the memoryless model coefficients. 

The \ac{GMP} model improves the test \ac{NMSE} over the memoryless model at every operating point (Table~\ref{tab:model_params}), by 1.4\,dB for the GaN PA (and by 1.3--1.8\,dB at all 11 drive levels) and by up to 4.6\,dB for the TLPA at 15\,GHz. 

The \ac{GMP} may also have a higher nonlinear order than the memoryless model, and part of this improvement can come from the order. Table~\ref{tab:same_order} separates the two variables: it compares the memoryless model, the best \ac{MP} ($B=0$) and the best \ac{GMP}, each selected with the same nonlinear order. At equal order, the memory terms alone give most of the improvement, the \ac{MP} improvement over the memoryless case is $4.1$\,dB for the TLPA at 15\,GHz/200\,MHz and $1.0$\,dB for the GaN PA). The \ac{GMP} cross terms add at most 0.4\,dB. The \ac{MP} therefore gives most of the memory gain with 14 to 42 coefficients, against 14 to 64 for the \ac{GMP}.

\begin{table}[t]
  \caption{Modelling results at fixed nonlinear order, with and without memory. For the memoryless (ML), \ac{MP} ($B=0$) and \ac{GMP} models, we list the number of coefficients / test \ac{NMSE} [dB]. Note $f_c$ in GHz, \ac{BW} in MHz.}
  \label{tab:same_order}
  \centering
  \footnotesize
  \setlength{\tabcolsep}{3pt}
  \begin{tabular}{llcccc}
    \hline
    \ac{DUT} & $f_c$/\ac{BW} & Order & ML & \ac{MP} & \ac{GMP} \\
    \hline
\inputrows{tables/tab_same_order.tex}
    \hline
  \end{tabular}
\end{table}

\begin{figure}[htbp]
  \centering
  \includegraphics[width=\linewidth]{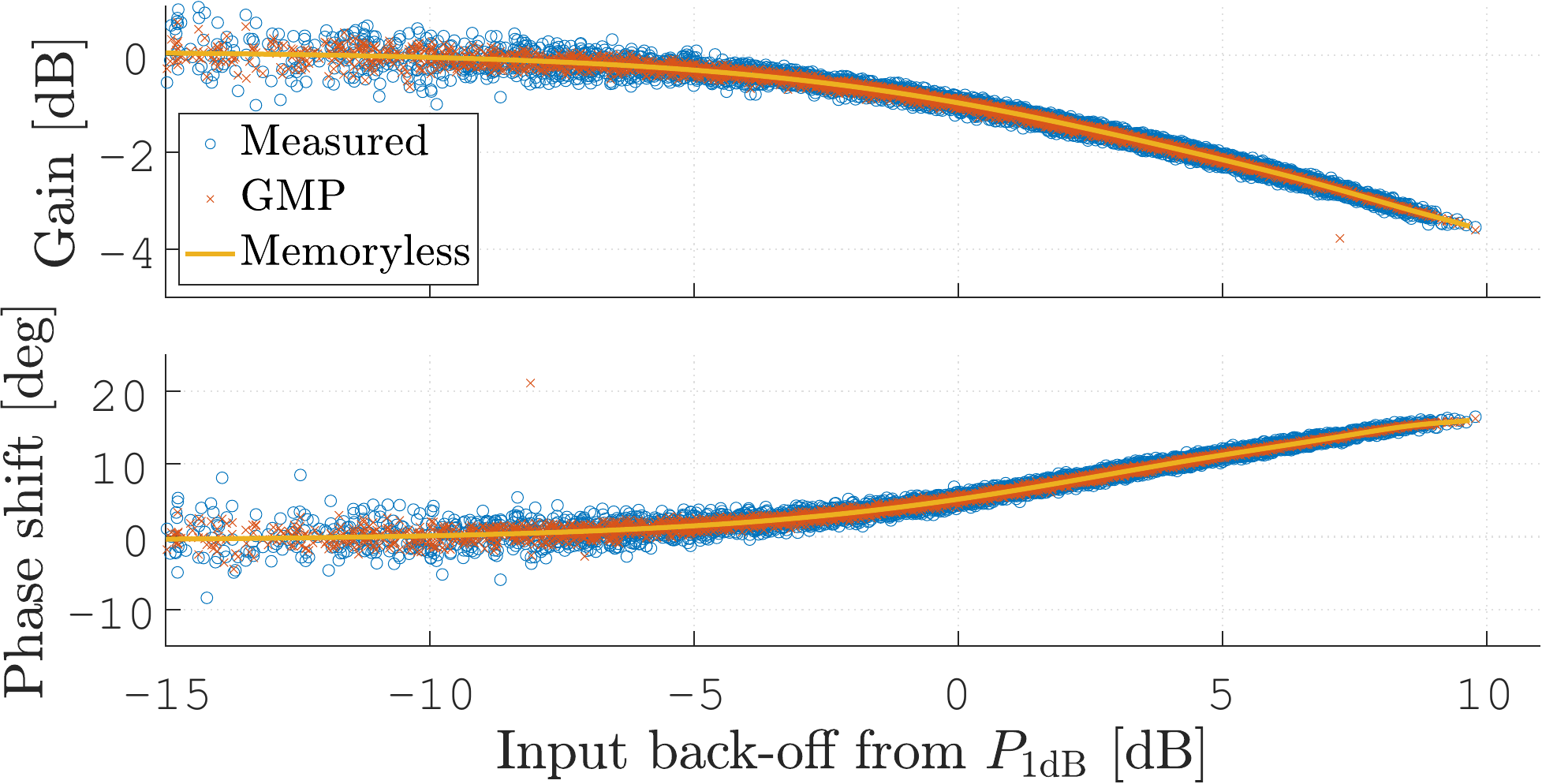}
  \caption{Memoryless vs.\ \ac{GMP} model: normalised AM/AM (top) and AM/PM (bottom), relative to the small-signal gain and phase, vs.\ input back-off from $P_\mathrm{1dB}$, for the GaN PA at the $-20$\,dBm drive level.}
  \label{fig:gmp_vs_ml}
  \vspace{-3mm}
\end{figure}

Fig.~\ref{fig:gmp_vs_ml} shows both the memoryless and \ac{GMP} models for the GaN PA at the $-20$\,dBm drive level, with the gain $|y|/|x|$ and phase $\angle y-\angle x$, compared to the measurement samples. The memoryless polynomial gives the mean gain and phase at each input amplitude. The \ac{GMP} output scatters around it, but less than the measurement, whose remaining spread neither model predicts. The \ac{GMP} needs 64 coefficients instead of 6 to gain 1.4\,dB of modelling accuracy. We conclude that if only aggregate distortion matters, such as the statistical average error or the sensing sidelobe floor (Section~\ref{sec:isac}), the memoryless model is the more computationally efficient option.

\begin{table*}[t]
  \centering
  \caption{Extracted memoryless polynomial coefficients (Eqn.~\eqref{eq:pa_poly}) for FR3 PA behavioral modelling, and link- or system-level simulation.\\ For details on use and models with memory, we refer to Section~\ref{sec:coefficients}.}
  \label{tab:coefficients}
  \centering
  \footnotesize
  \setlength{\tabcolsep}{3pt}
  \begin{tabular}{lccccc}
    \hline
     & GaN PA & TLPA & TLPA & TLPA & GaAs LNA \\
     & 8\,GHz, 100\,MHz & 7.225\,GHz, 200\,MHz & 15\,GHz, 100\,MHz & 15\,GHz, 200\,MHz & 15\,GHz, 100\,MHz \\
    \hline
\inputrows{tables/tab_coefficients.tex}
    \hline
  \end{tabular}
\end{table*}

\subsection{Impact of Modulation and Bandwidth}\label{sec:bw_comparison}
\begin{figure}[htbp]
  \centering
  \includegraphics[width=\linewidth]{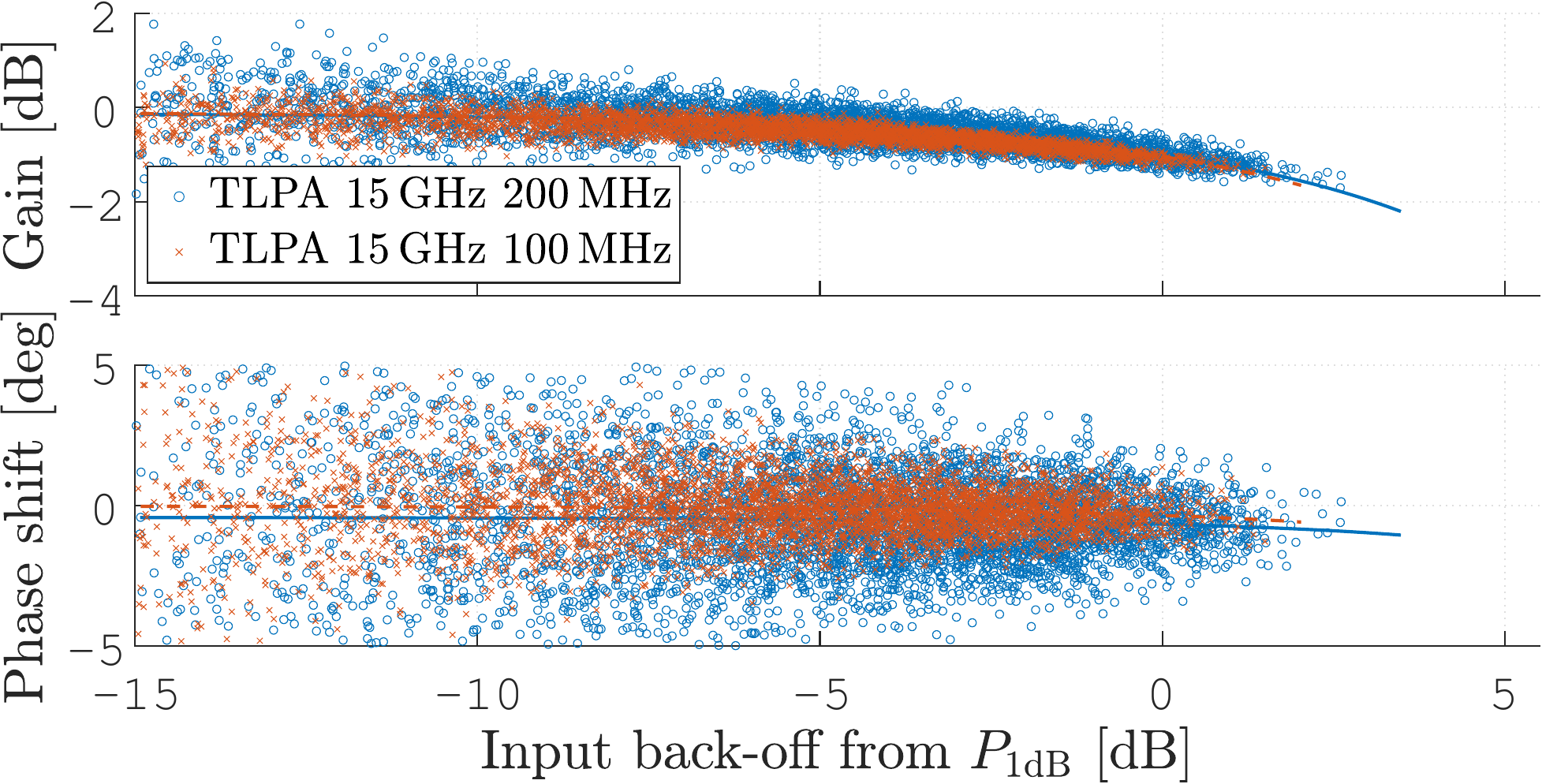}
  \caption{Normalised AM/AM (top) and AM/PM (bottom), relative to the small-signal gain and phase, vs.\ input back-off from $P_\mathrm{1dB}$: measured scatter and fitted polynomial for TLPA at 15\,GHz, comparing 100\,MHz and 200\,MHz signal bandwidths at $-20$\,dBm input power.}
  \label{fig:amam_ampm_bw}
  \vspace{-3mm}
\end{figure}

At 15\,GHz, the TLPA was measured with 100\,MHz 64-\ac{QAM} and 200\,MHz 256-\ac{QAM} signals at the same input power. In Fig.~\ref{fig:amam_ampm_bw}, we show the normalised AM/AM and the AM/PM for both measurements. The compression trend is the same for both, only the spread around it widens at 200\,MHz. We found significant linear memory contributions in the TLPA datasets, which might indicate a matching network linear response. The test \ac{NMSE} of the 3rd-order polynomial rises from $-31.8$ to $-28.1$\,dB with the higher bandwidth. Because bandwidth and modulation order change together, this gap reflects their joint effect.

\subsection{ISAC Evaluation}\label{sec:isac}
\begin{figure}[htbp]
  \centering
  \includegraphics[width=\linewidth]{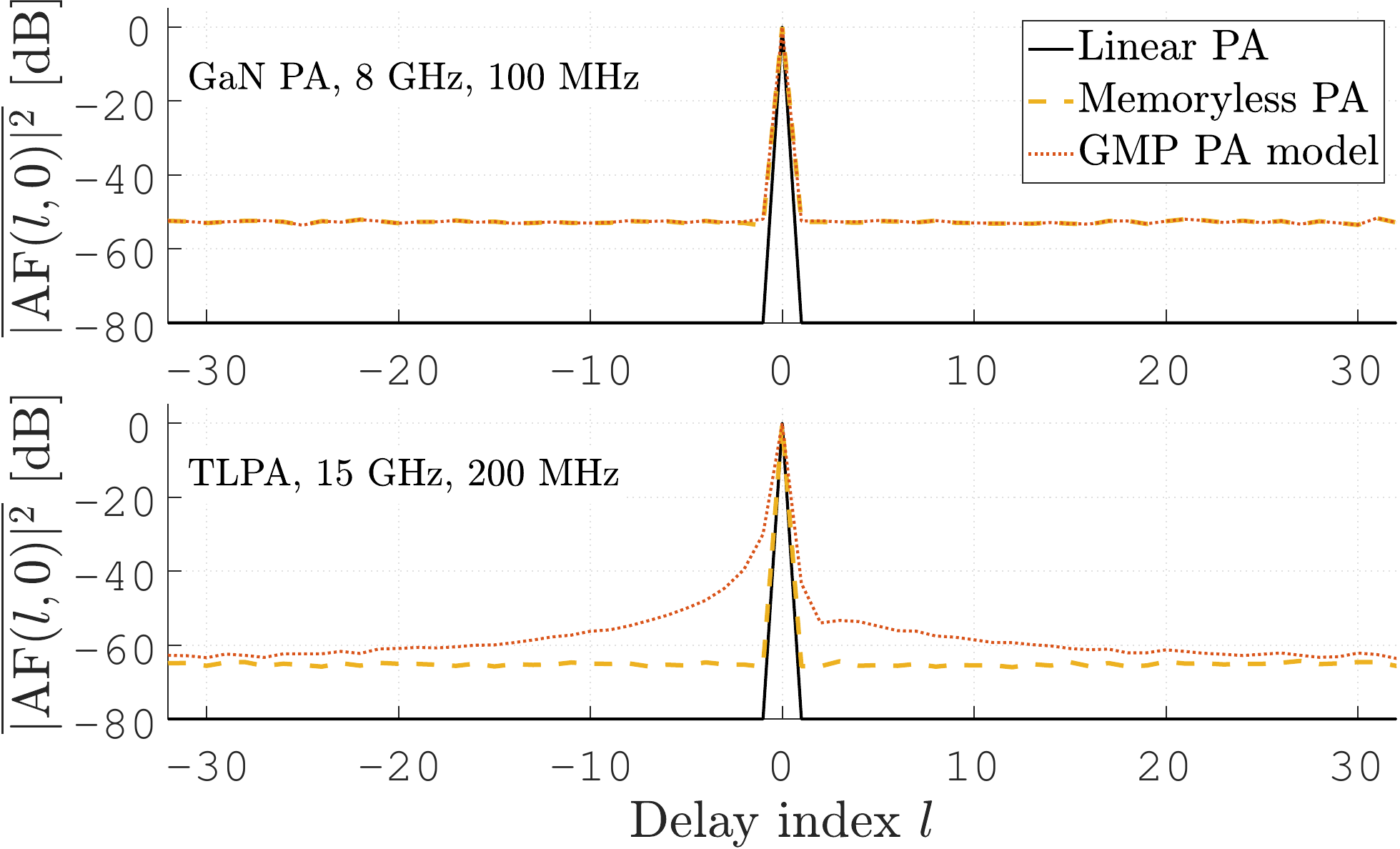}
  \caption{Zero-Doppler cut of the per-symbol matched-filter ambiguity function of a 16-PSK CP-OFDM frame through a linear PA and the memoryless and \ac{GMP} models, evaluated at integer delay bins as in~\cite{gourar2025pa_isac}. Top: GaN PA (highest drive level). Bottom: TLPA at 15\,GHz/200\,MHz.}
  \label{fig:isac}
  \vspace{-3mm}
\end{figure}

To show how the measured PA nonlinearity affects sensing, we follow the OFDM ambiguity-function analysis of~\cite{gourar2025pa_isac}, replacing its soft-limiter model with our memoryless and \ac{GMP} PA models at the highest drive level. We consider a monostatic scenario with a CP-OFDM signal using $N=2048$ subcarriers over 100\,MHz \ac{BW} and $M=128$ symbols. We use 16-PSK modulation since constant-modulus subcarriers have zero matched-filter data sidelobes, so all sidelobe energy comes from the PA~\cite{gourar2025pa_isac}. The sample rate equals the measured one, so the \ac{GMP} model taps keep their physical meaning. The signal is scaled so that its peaks span the measured input amplitude range and is clipped at the largest fitted amplitude (0.01\,\% of the samples), which keeps the models within their fitted range. Because of the higher \ac{PAPR} of OFDM, its RMS level is then about 3\,dB below that of the measured signal. The PA outputs are normalised by their small-signal gain. Fig.~\ref{fig:isac} shows the zero-Doppler cut of the per-symbol matched-filter ambiguity function at integer delay bins, averaged over the $M$ symbols~\cite[eq.~(20)]{gourar2025pa_isac}.

We note from Fig.~\ref{fig:isac} that the linear amplification has a sidelobe floor of $-88$\,dB, set by the clipping filtering. For the GaN PA, compression raises this floor to $-53$\,dB, with the same effect for both the memoryless and \ac{GMP} models. The less compressed TLPA (15\,GHz/200\,MHz) reaches a sidelobe floor of $-65$\,dB with a memoryless model. The \ac{GMP} model of the TLPA \ac{DUT} adds a shoulder within about 15 delay bins of the mainlobe, but the floor further out is the same. The shoulder comes from the linear memory effect of the TLPA, an in-band ripple of 0.7\,dB and $6^\circ$ over 200\,MHz, which is deterministic and can be equalised. The floor is set by the nonlinear distortion, which the memoryless model also captures.


\subsection{Model Coefficients}\label{sec:coefficients}
Table~\ref{tab:coefficients} lists the complex coefficients $a_{2k+1}$ of the memoryless polynomial model~\eqref{eq:pa_poly} for the measured operating points summarised in Table~\ref{tab:model_params}, for use in system-level simulation.
As in the 3GPP baseline, the coefficients apply to an input normalised to unit peak amplitude, $|x(n)|\le1$. A user applying them to a different waveform should first scale its envelope to the intended operating point, which is the desired back-off from $P_\mathrm{1dB}$ (see Section~\ref{sec:model_eval}).
The \ac{MP} and \ac{GMP} coefficients are published with the measurement data~\cite{dataset_306iim}.

\section{Conclusion}\label{sec:conclusion}
This paper presented measurement-based PA behavioral models for FR3: a GaN PA at 8\,GHz, driven up to 5.2\,dB compression over 11 drive levels, and a wideband PA at 7.225 and 15\,GHz with 100\,MHz and 200\,MHz \ac{BW}.  To the best of our knowledge, these measurements form the first publicly released FR3 PA dataset.
A 3rd- to 5th-order memoryless polynomial suffices to model the wideband PA, while the GaN PA exhibits soft compression and is best modelled by an 11th-order polynomial.
The extracted models follow the 2.1\,GHz GaN baseline of TR~38.803 up to about 4\,dB beyond $P_\mathrm{1dB}$. The GaN PA shows pronounced nonlinear phase distortion, whereas the TLPA remains within $1^\circ$.
The \ac{GMP} model, which includes memory terms, improves the test \ac{NMSE} at every operating point, by 1.3\,dB to 4.6\,dB. At the same nonlinear order, a memory polynomial without cross terms already provides most of this gain, with fewer coefficients.
In an OFDM ISAC ambiguity-function evaluation, the compressed GaN PA raises the sidelobe floor by 35\,dB. The memoryless and \ac{GMP} models give the same floor, while the \ac{GMP} additionally shows close-in sidelobes from linear, equalisable memory. The memoryless coefficients of Table~\ref{tab:coefficients} suffice to predict the floor, and can be used in link- and system-level simulations.
These findings describe single units of three devices, measured with single-carrier 64- and 256-\ac{QAM} signals. Future work will further investigate the impact of GaN PA memory effects on sensing and provide measurement data as additional FR3 bands are investigated in standardization.

\ifaccepted
\section*{Acknowledgment}
This work was supported by the Smart Networks and Services Joint Undertaking (SNS JU) under the European Union's Horizon Europe research and innovation program under Grant Agreement No 101192369 (6G-MIRAI). Views and opinions expressed are however those of the authors only and do not necessarily reflect those of the European Union or the SNS JU. Neither the European Union nor the granting authority can be held responsible for them. The work of R. Martinez Alonso and Z. Cui is supported by the Research Foundation--Flanders (FWO) through Senior Postdoctoral Fellowships under Grants 1211926N and 12AFN26N, respectively.
\fi

\bibliography{references}
\bibliographystyle{IEEEtran}

\end{document}